\documentclass[11pt,aps,prd,preprint,preprintnumbers,showpacs,superscriptaddress,floatfix]{revtex4-1}
\usepackage{graphicx}
\usepackage{epsfig}
\usepackage{dcolumn}
\usepackage{bm}
\usepackage{subfigure}
\usepackage{amsmath}
\usepackage{amssymb}
\usepackage{slashed}
\usepackage[T1]{fontenc}
\usepackage[latin1]{inputenc}
\usepackage{caption}
\usepackage{multirow}
\usepackage{setspace}
\usepackage{xcolor}
\usepackage[colorlinks,
            linkbordercolor={0 0 0},
            pdfborder={0 0 0},
            linkcolor=blue,
            citecolor=blue,
            urlcolor=blue,
            breaklinks=true]{hyperref}

\allowdisplaybreaks

\newlength{\x}
\newlength{\y}
\newlength{\z}
\urldef\milcurl\url{http://www.physics.utah.edu/~detar/milc/}

\phantomsection

\begin{document}
\preprint{IMSc/2023/01}

\title{Parametric resonance in abelian and non-abelian gauge fields via  space-time oscillations}

\author{Shreyansh S. Dave}
\email{shreyansh.dave@tifr.res.in}          
\affiliation{Department of Nuclear and Atomic Physics, Tata Institute of Fundamental Research, Mumbai 400005, India}

\author{Sanatan Digal}
\email{digal@imsc.res.in}           
\affiliation{The Institute of Mathematical Sciences, Chennai 600113, India}
\affiliation{Homi Bhabha National Institute, Training School Complex,
Anushakti Nagar, Mumbai 400094, India}

\author{Vinod Mamale}
\email{mvinod@imsc.res.in}
\affiliation{The Institute of Mathematical Sciences, Chennai 600113, India}
\affiliation{Homi Bhabha National Institute, Training School Complex,
Anushakti Nagar, Mumbai 400094, India}

\begin{abstract}

We study the evolution of abelian $U(1)$ electromagnetic as well as non-abelian $SU(2)$ gauge fields, in the presence of space-time oscillations. Analysis of the time evolution of abelian gauge fields shows the presence of parametric resonance in spatial modes. A similar analysis in the case of non-abelian gauge fields, in the linear approximation, shows the presence of the same resonant spatial modes. The resonant modes induce large fluctuations in physical observables including those that break the $CP-$symmetry. We also carry out time evolution of small random fluctuations of the gauge fields, using numerical simulations in $2+1$ and $3+1$ dimensions. These simulations help to study non-linear effects in the case of non-abelian gauge theories. Our results show that there is an increase in energy density with the coupling, at late times.  These results suggest that gravitational waves may excite non-abelian gauge fields more efficiently than electromagnetic fields. Also, gravitational waves in the early Universe and from the merger of neutron stars, black holes etc. may enhance $CP-$violation and generate an imbalance in chiral charge distributions, magnetic fields etc.

\end{abstract}


\maketitle

\section{Introduction} 
\label{sec:intro} 

The phenomenon of parametric resonance(PR), in particle mechanics~\cite{Lan:1976}, arises when parameters of the system are oscillatory. This phenomenon finds applications in diverse systems, ranging from parametric oscillators
to particle productions in the early Universe~\cite{Ide:1995,Wu:1986zz,Figueroa:2016wxr,Araujo:2021}. 
In the context of the early Universe, entropy is generated via the mechanism of
parametric amplification, which subsequently leads to particle production~\cite{Hu:1986jd,Hu:1986jj,Kandrup:1988sg}. During the
preheating stage of the post-inflationary period, an oscillatory super massive field leads to explosive production of particles~\cite{Traschen:1990sw,Kofman:1994rk,Kofman:1997yn,Zibin:2000uw,Jaeckel:2021xyo,Joana:2022uwc}. In the context of heavy-ion 
collisions, the parametric amplification of pion field induced by oscillatory sigma meson field has been explored~\cite{Boyanovsky:1994me}. 
It has been found that an oscillatory free energy~\cite{Digal:1999ub}, oscillatory trapping potential~\cite{oscBEC1,oscBEC2,resoBEC1} etc. also
give rise to parametric resonance. In the presence of gravitational waves (GWs), parametric resonance is observed in neutrino spin/flavor oscillations in presence of gravitational waves\cite{Dvornikov:2019xok,Losada:2022uvr}. It is hypothesised that coherently oscillating axion condensate will exhibit parametric resonance in an electromagnetic field coupled through the Chern Simons term and which can eventually lead to radio wave emission~\cite{Yoshida:2017ehj,Hertzberg:2018zte,Arza:2020eik}.

Recently, it has been shown that transversely polarised (GWs), i.e., sustained mono-chromatic space-time oscillations, 
induce large fluctuations in a superfluid condensate~\cite{Dave:2019iwr}. These large fluctuations subsequently 
decay to vortex-anti-vortex pairs, without having the system go through a phase transition. Detailed numerical analysis, of the momentum modes of the fields in the linear regime, showed that they grow exponentially. In this regime, the analysis of the time evolution equations of the field modes, which resemble that of a parametric oscillator, confirms that the underlying phenomenon is the parametric resonance. The resonant growth of field modes is found to be significant also in the case of transient and non-mono-chromatic GWs, i.e., those generated during merger 
events of neutron stars and/or black holes. In ref~~\cite{Dave:2021lcv}, a GW pulse modelled on the merger event GW150914, detected by the LIGO experiment, leads to an exponential growth in the fluctuations in a fuzzy dark matter.  

The source of parametric resonance in Ref.\cite{Dave:2019iwr,Dave:2021lcv} is oscillatory gradients in the field 
equations, which arise from the coupling between the classical fields and the space-time oscillations. The coupling of space-time oscillations to 
gauge fields being similar, we expect the electromagnetic, as well as the non-abelian gauge fields, will also undergo parametric resonance. 
We mention here that the interaction between electromagnetic waves(EMWs) and GWs has been extensively studied since Einstein's general 
theory of relativity predicted their existence~\cite{Plebanski:1959ff,Griffiths:1975zm,cooperstock,Barrabes:2010tr,Morozov:2021,Barreto:2017shm,Mishima:2022xrq,Sotani:2014fia,Jones:2017dzt,Wang:2018yzu,Jones:2019nsp,Patel:2021cat,Brandenberger:2022xbu}. GWs affect various properties of EMWs and vice-versa. The Polarization of EMWs gets rotated through scattering with GWs ~\cite{Plebanski:1959ff}. The GWs and EMWs are known to induce fluctuations in each other~  \cite{cooperstock,Barrabes:2010tr,Morozov:2021}. Further enhancement in the fluctuations of EMWs is observed when non-linear interactions are included~\cite{Barreto:2017shm,Mishima:2022xrq}. There are effects of GWs on EMWs which lead to interesting physical consequences \cite{Sotani:2014fia,Jones:2017dzt,Wang:2018yzu,Jones:2019nsp}. For example, in the case of the core collapse of a pre-supernova star, the production of electromagnetic waves can lead to fusion outside of the iron core~\cite{Jones:2019nsp}. Recently, the 
response of electromagnetic fields to GWs has been studied by evolving them in the presence of oscillating background the metric. A small oscillating background on top of flat spacetime causes deviation in direction of EMW when it is perpendicular to GW's direction of propagation~\cite{Patel:2021cat}. The simultaneous, electromagnetic and gravitational wave signal detection of binary neutron star merger event GW170817 will shed more light on the interplay between  GWs and EMWs ~\cite{Abbott:2017}. In De-Sitter space, the perturbations of electromagnetic fields grow with conformal time. 
The space-time oscillations, around a flat space-time background, drive the parametric resonance of electromagnetic fields, as a result of which the GWs
get damped while passing through a medium~\cite{Brandenberger:2022xbu}. Quantum corrections to electrodynamics, due to inflationary gravitons in De-Sitter space, has also been studied previously~\cite{Leonard:2013xsa,Glavan:2013jca,Wang:2014tza,Miao:2018bol}.

In this work, we study the parametric resonance in electromagnetic ($U(1)$) as well as non-abelian ($SU(2)$) gauge fields in the presence of space-time oscillations, using analytical methods and numerical simulations. As in ~\cite{Dave:2019iwr,Dave:2021lcv}, the space-time oscillations are considered to be monochromatic waves, which modify the classical field equations. In the case of $U(1)$ fields, the modes undergoing parametric resonance are obtained from the field equations in momentum space. For $SU(2)$, the mode equations are analysed in the linear approximation as different modes effectively decouple. The mode equation for each color field becomes identical to that of $U(1)$ gauge fields'. In the numerical simulations, initial field configurations are dynamically evolved in real-time, in $2+1$ and $3+1$ dimensions. The initial field configurations are such that the energy of the system entirely comes from small fluctuations. These simulations are crucial to compare the dynamics of fluctuations in $SU(2)$ and $U(1)$ gauge theories. Further, the numerical simulations are useful in $U(1)$, even though the theory is linear, to evolve field configurations with special boundary conditions. 
\par As expected, our numerical results show that in the linear regime, the dynamics of fluctuations in $U(1)$ and $SU(2)$ gauge theory are similar. The field modes undergo exponential growth due to parametric resonance. In the case of the $SU(2)$ gauge theory, initially in the non-linear regime, the energy density starts to decrease. Subsequently, as the fluctuations undergo sustained parametric resonance, a larger growth is observed compared to the case of vanishing gauge coupling. The details of decay and eventual enhancement depend on the initial configuration, i.e., on the distribution of modes.  We mention here that the dynamics of gauge fields is similar to that of fluctuations in constant chromo-magnetic fields, the so-called Nielsen-Olesen(N-O) instability\cite{Nielsen:1978rm,Nielsen:1978nk}. In the latter case, along with exponentially growing modes, overdamping modes are also present. The overdamping modes' effects are largest initially and decrease with time~\cite{Bazak:2021xay}. Large chromo-electric fields are also known to cause the N-O instability\cite{Chang:1979tg}. Previous studies in presence of oscillatory chromo-magnetic fields, show there is a subdominant growth due to parametric resonance along with dominant N-O instability~\cite{Berges:2011sb,Tsutsui:2014rqa,Tsutsui:2015ule}. The above non-linear effects suggest that GWs will excite $SU(2)$ gauge fields more efficiently than $U(1)$ electromagnetic fields.

The exponential growth of modes lead to large fluctuations in physical observables such as energy density, ${\bf E}\cdot{\bf B}$ in $U(1)$ and $tr(F\tilde{F})$ in $SU(2)$ gauge theories etc. Our numerical simulations show that even if 
${\bf E}\cdot{\bf B}$ and $tr(F\tilde{F})$ are zero initially, they develop non-zero values and grow, due to the presence of multiple resonant modes. Since ${\bf E}\cdot{\bf B}$ and $tr(F\tilde{F})$ are related to the divergence of chiral current, they can contribute to chiral magnetic effects, enhance instanton transitions, production of particles etc.  It will be interesting to consider these processes in the context of cosmology. We mention here that, the resonant excitation of fields in QCD 
or electroweak theory, will not be possible from the gravitational waves from merger events. This is because of the presence of the mass scales, such as $\Lambda_{QCD}$. However, phase transitions in the early Universe may generate gravitational waves which can induce parametric resonance in QCD~\cite{Fenu:2009qf,Bodas:2022urf,Hindmarsh:2020hop}.

The paper is organised as follows. In section II we obtain the parametric resonant modes of the fields due to space-time oscillations. In 
section III, we present the results of our numerical simulations of $U(1)$ electromagnetic fields and $SU(2)$ non-abelian gauge fields.
The conclusion and discussions are presented in section IV.

\section{ Parametric resonance in the linear approximations}

Space-time oscillations are taken into account by considering a flat metric with time-dependent terms, i.e., 
$g_{\mu\nu} = diag(-1,1-\epsilon\sin(\omega(t-z)),1+\epsilon\sin(\omega(t-z)),1)$, with $\epsilon < 1$~\cite{Schutz,Bishop:2016lgv}. $\omega$ is the frequency
of the monochromatic plane GW, propagating along the $z-$direction. For simplicity, we ignore the back
reaction on the metric due to the gauge fields. The Lagrangian density for $SU(2)$ gauge fields is given by,
\begin{equation}
L=-\frac{1}{4}g^{\alpha\mu}g^{\beta\nu}F^a_{\alpha\beta}F^a_{\mu\nu}.
\label{lgrn}
\end{equation}
The field strength tensor defined as $F^a_{\mu\nu}=\nabla_{\mu}A^a_{\nu}-\nabla_{\nu}A^a_{\mu} + g\epsilon_{abc}A^b_{\mu}A^c_{\nu}$, where $A^a_\mu$ denotes the four-vector gauge field with color $a$. $\epsilon_{abc}$ is the Levi-Civita symbol in color space, $a,b,c=1,2,3$. $g$ is the gauge coupling constant and is taken to be one in our study, otherwise mentioned. The covariant derivative is defined as $\nabla_\mu A^{a\nu}=\partial_\mu A^{a\nu} + \Gamma^\nu_{\mu \alpha}A^{a\alpha}$, where $\Gamma^\nu_{\mu \alpha}$ are the Christoffel symbols. In the Lorentz gauge, $\nabla_{\eta}A^{a\eta}=0$, the Euler-Lagrange equation for the gauge field $A^{a\nu}$, is given by,
\begin{eqnarray}
\nabla_{\mu}\nabla^{\mu}A^{a\nu}-g^{\lambda\nu}R_{\sigma\lambda}A^{a\sigma}+gf^{abc}A^{b\mu}\nabla_{\mu}A^{c\nu}\nonumber\\+gf^{abc}A^{b}_{\mu}\left(\nabla^{\mu}A^{c\nu}-\nabla^{\nu}A^{c\mu}\right)+g^2\left(A^{b\nu}A^b_\mu A^{a\mu}-A^{b\mu} A^{a\nu}A^{b}_\mu\right)=0
\label{eom}.
\end{eqnarray}
Here, $R_{\sigma\lambda}$ is Ricci tensor. Analytically the resonant modes can be obtained by considering linear approximation, which can also be done by setting $g=0$. In this approximation, different field modes, as well as colors, effectively decouple and their time evolution resembles that of a parametric oscillator. Note that, for $U(1)$, the Lagrangian density and field equations can be obtained by setting $g=0$ above and dropping the color indices. Thus the gauge fields in $SU(2)$ with $g=0$ and $U(1)$ gauge fields will undergo the same dynamical evolution. The color indices, therefore, do not play any essential role. The field equations for $U(1)$ and for $SU(2)$ in linear approximation take the following form,
\begin{eqnarray}
-A_{0,tt}+A_{0,zz}+\frac{A_{0,xx}}{1-h}+\frac{A_{0,yy}}{1+h}+\frac{2hh_tA_{0,t}-2hh_zA_{0,z}}{2(1-h^2)}+R(t)A_0-R(t)A_3 = 0,~~~~~~~
\label{fld0}
\end{eqnarray}
\begin{eqnarray}
\frac{A_{1,xx}}{1-h}+\frac{A_{1,yy}}{1+h}+\frac{2hh_tA_{1,t}-2hh_zA_{1,z}}{2(1-h^2)}-A_{1,tt}+A_{1,zz}=0,
\label{fld1}
\end{eqnarray}
\begin{eqnarray}
\frac{A_{2,xx}}{1-h}+\frac{A_{2,yy}}{1+h}+\frac{2hh_tA_{2,t}-2hh_zA_{2,z}}{2(1-h^2)}-A_{2,tt}+A_{2,zz}=0,
\label{fld2}
\end{eqnarray}
and,
\begin{eqnarray}
-A_{3,tt}+A_{3,zz}+\frac{A_{3,xx}}{1-h}+\frac{A_{3,yy}}{1+h}+\frac{2hh_tA_{3,t}-2hh_zA_{3,z}}{2(1-h^2)} + R(t) A_0 - R(t)A_3=0,~~~~~~~
\label{fld3}
\end{eqnarray}
for the fields $A_0,~A_1,~A_2,$ and $~A_3$ respectively.  The second (third) subscript represents first (second) derivatives. $h\equiv h(z,t)=\epsilon sin(\omega(t-z))$ and it's derivatives $w.r.t$ $z$ and $t$ are represented by the corresponding subscripts. 
\begin{equation}
R(t)\equiv R^0_0(t) = \left[\frac{h_th_t}{4(1-h)^2}+\frac{h_th_t}{4(1+h)^2}+\frac{hh_{tt}}{1-h^2}\right]
\end{equation}
is the Ricci tensor $R^\mu_{\nu}$ with $\mu=\nu=0$. 
\subsection{The time evolution of the momentum modes}
The evolution equations for the momentum modes, $a_\mu({\bf k},t)$, are obtained by writing $A_\mu({\bf x},t)$  as
\begin{equation}
A_\mu({\bf x},t)={1 \over V} \int d^3k a_{\mu}({\bf{k}},t)e^{i{\bf{k\cdot}\bf{x}}}.
\end{equation}
Since there are no spatial oscillations of the metric along the $z-$direction, it is expected that there will not be any resonant growth of $z-$component
momentum modes. Therefore we consider, ${\bf x}=(x,y)$ and ${\bf k}=(k_x,k_y)$ and set $z=0$. The time evolution equations, for the momentum modes, are given by,
\begin{equation}
-C(t)\dot{a}_{0}({\bf{k}},t)+\ddot{a}_{0}({\bf k},t)+\left(k^2_xf(t)+k^2_yf(-t)-R^0_0(t)\right)a_{0}({\bf k},t)+R^0_0(t)a_{3}({\bf k},t)=0,
\label{eq0}
\end{equation}
\begin{equation}
-C(t)\dot{a}_{3}({\bf{k}},t)+\ddot{a}_{3}({\bf k},t)+\left(k^2_xf(t)+k^2_yf(-t)+R^0_0(t)\right)a_{3}({\bf k},t)-R^0_0(t)a_{0}({\bf k},t)=0,
\label{eq3}
\end{equation}
\begin{equation}
-C(t)\dot{a}_{i}({\bf k},t)+\ddot{a}_{i}({\bf k},t)+\left(k^2_xf(t)+k^2_yf(-t)\right){a}_{i}({\bf k},t)=0,~i=1,2.
\label{eqij}
\end{equation}
$f(t)=1/(1-\epsilon\sin(\omega t))$, $C(t)=\epsilon^2\omega f(t)f(-t)sin(2\omega t)/2$, and ``$\cdot$'' represents derivative w.r.t time.
Note that Eq.$(\ref{eq0}-\ref{eqij})$, are valid for $SU(2)$ only in the linear approximation, i.e. when the amplitudes of the field modes are small. 
In all the above equations, damping terms oscillate with frequency $2\omega$. In Eq.\ref{eq0} and Eq.\ref{eq3}, the external force
in the form of Ricci tensor also oscillates with frequency $\omega$. The above mode equations resemble that of a parametric
oscillator with an oscillatory damping term. It is well known that these oscillatory terms drive paramagnetic resonance in the following momentum modes,
\begin{equation}
k_{x,y} = {n\omega\over 2},
\label{reso}
\end{equation}
where $n=1,2,3....$. In the linear approximation in $SU(2)$, which is
valid when the fluctuations are small, the different modes of the gauge fields decouple. So initially the modes which satisfy Eq.\ref{reso} will
undergo resonance.  Once the field modes grow beyond the linear regime, the interaction between gauge fields of different colors will be necessary. 
In this situation, evolution can only be studied by using numerical simulations.
\subsection{Effect of spacetime oscillations on $F\tilde{F}$}
The resonant growth of modes induces large fluctuations in the field strength tensor. In the following, we analyse the time evolution of $F_{\mu\nu}\tilde{F}^{\mu\nu}$ $({\bf E}\cdot{\bf B})$ in $U(1)$  and $F^a_{\mu\nu}\tilde{F}^{a\mu\nu}$ $(tr(F\tilde{F}))$, in $SU(2)$ gauge theories, using the above mode equations. In the linear approximation, the color indices can be ignored. The electric fields ${\bf E}=\vec{E}$ and magnetic fields ${\bf B}=\vec{B}$ can be written in terms of the momentum modes, as
\begin{eqnarray}
\vec{E}(\vec{x},t)&=&\frac{1}{V}\int  d^2k \left[-\left\{ i k_x a_0({\bf k},t)+\dot{a}_1({\bf k},t)\right\}\hat{i}-\left\{i k_y{a_0({\bf k},t)}+\dot{a}_2({\bf k},t)\right\}\hat{j}-\dot{a}_3({\bf k},t)\hat{k}\right]e^{i \vec{k}\cdot\vec{x}}\nonumber\\
\vec{B}(\vec{x},t)&=&\frac{1}{V}\int d^2k \left[ i k_y {a_3({\bf k},t)}\hat{i}-i k_x {a_3({\bf k},t)}\hat{j}+\left\{i k_x {a_2({\bf k},t)}-i k_y {a_1({\bf k},t)}\right\}\hat{k}\right]e^{i \vec{k}\cdot\vec{x}}
\end{eqnarray}
The condition $\vec{E}\cdot\vec{B}=0$, leads to the following equation,
\begin{eqnarray}
i k_x\left\{\dot{a}_2({\bf k},t)a_3({\bf k},t)-a_2({\bf k},t)\dot{a}_3({\bf k},t)\right\} + i k_y\left\{a_1({\bf k},t)\dot{a}_3({\bf k},t)-\dot{a}_1({\bf k},t){a_3}({\bf k},t)\right\}=0
\end{eqnarray}
in terms of the gauge field modes.  In order that ${\bf E}\cdot{\bf B}$ remains zero at later times, the following equation must be 
satisfied,
\begin{equation}
{\dot{a}_i({\bf k},t)\over a_i({\bf k},t)}={\dot{a}_3({\bf k},t)\over a_3({\bf k},t)},~i=1,2.
\end{equation}
This condition requires that the modes, $a_1({\bf k},t), a_2({\bf k},t)$ and $a_3({\bf k},t)$ grow in sync, which is not ensured by evolution equations \ref{eq3},\ref{eqij}. Even with ${\bf E}\cdot{\bf B}=0$ initially, 
large values can be generated subsequently via parametric resonance. In the numerical simulations, we also consider the time evolution of gauge fields starting with zero ${\bf E}\cdot{\bf B}$, for both abelian and non-abelian cases. 

\section{Numerical Simulations}

In this section, we describe numerical simulations of the field equations and the results, for $U(1)$ and $SU(2)$ gauge theories. In $3+1$ dimensions, space-time oscillations propagating in the $z-$direction result in adjacent $z-$planes oscillating with a finite phase difference. As a consequence, fields in different $z-$planes evolve differently and thus generate $k_z$ modes even if they are not present initially. Further in the case of $SU(2)$, $k_z$ modes will also get excited due to non-linear interactions. 
For simplicity, we assume the fields are uniform along the $z-$direction and carry out numerical simulations in $2+1$ dimensions.  
We also consider simulations in $3+1$ dimensions, to see the effect of $z-$modes on the overall resonant growth of the fluctuations. The results of $3+1$ studies are discussed at the end of section III-A.

In the simulations, the two-dimensional ($x,y$) plane of area $L^2$ is discretised as a $(N\times N)$ lattice. We consider $N=200$ for most of our simulations. The lattice constants are taken to be the same in both $x$ and $y$ directions, i.e., $\Delta x=\Delta y=L/N=0.01 \Lambda^{-1}$. The time is also discretised with time spacing $\Delta t$. Stability in the numerical evolution requires $\Delta t < \Delta x/\sqrt{2}$, so we take $\Delta t = 0.005\Lambda^{-1}$. Note that in $U(1)$ gauge theory there is no natural scale. The same is true for the $SU(2)$ gauge theory at the classical level. In this situation, we take the $\Lambda$ scale to be the same as the inverse of the time period of the space-time oscillations. For most of the simulations, we take $\omega=50\Lambda$ and $\epsilon=0.4$. For a few simulations, to check the effect of discretisation, we consider $\omega=16\pi \Lambda$ and $\Delta x < 0.01 \Lambda^{-1}$. This choice of frequency satisfies periodic boundary conditions and is suitable for studying discretization effects.

\par The field equations are discretised using a second-order leap-frog algorithm~\cite{Press:1997}. Since the field equations are second-order, the evolution requires fields at two different times initially, i.e., at $t=0$ and $\Delta t$. At both these times the gauge fields are taken to be the same, but at each lattice point on the two-dimensional lattice,  $A_\mu(n_1,n_2)$, are chosen randomly with uniform probability in the range $[-\gamma,\gamma]$ with $\gamma=0.005\Lambda$.  For some simulations with specific initial conditions for  ${\bf E\cdot B}$ studies, larger lattices are considered. This is useful to avoid boundary effects up to larger times. With the above choice of parameters, the initial distributions of ${\rm E}^2$  and  ${\rm B}^2$ are as small as $O(1)$ in the case of abelian gauge fields. Due to space-time oscillations ($\omega=50\Lambda$), fluctuations of the gauge fields grow exponentially in time. As a consequence, both the electric and magnetic fields also grow. 
\begin{figure}[ht]
	\begin{minipage}{0.45\textwidth}
		\includegraphics[width=\textwidth]{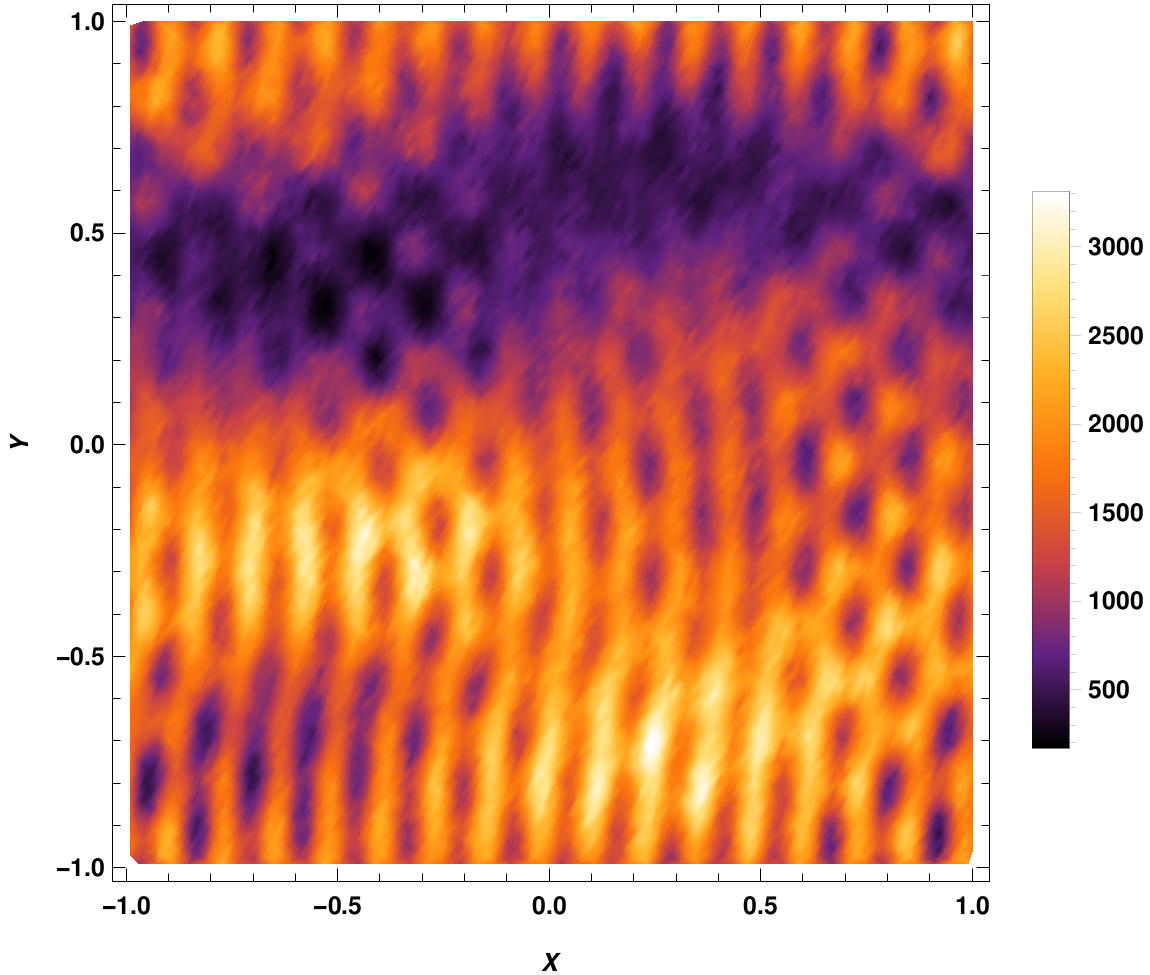}
		\caption{${\bf E}^2$ at t=4$\Lambda^{-1}$ for initial configuration with random fluctuations}
		\label{E2}
	\end{minipage}
	\begin{minipage}{0.45\textwidth}
		\includegraphics[width=\textwidth]{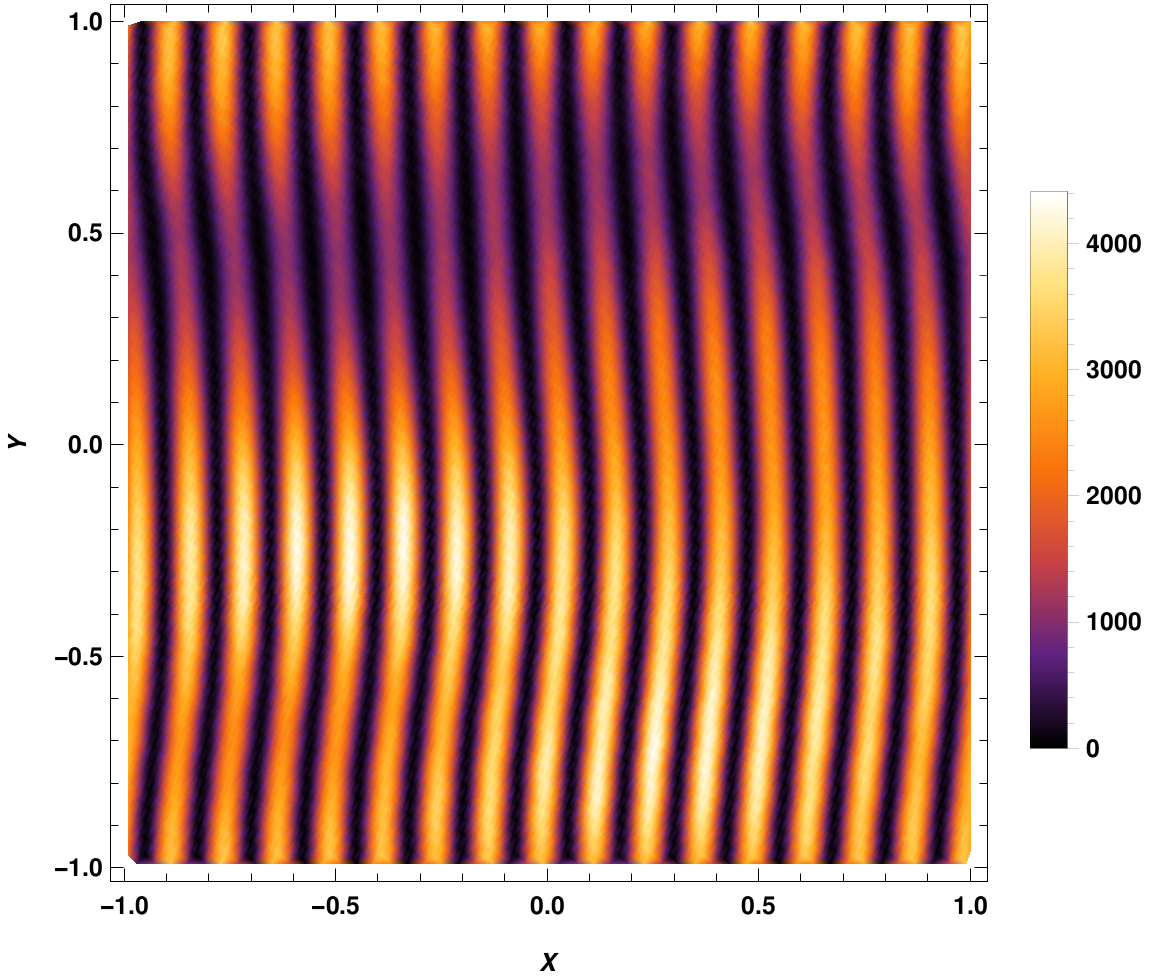}
		\caption{${\bf B}^2$ at t=4$\Lambda^{-1}$ for initial configuration with random fluctuations}
		\label{B2}
	\end{minipage}
\end{figure}
In Fig.\ref{E2} and \ref{B2}, $\rm{E}^2$ and ${\rm B}^2$ in units of $\Lambda^4$, are plotted at $t=4\Lambda^{-1}$. It can be seen that within time interval $4\Lambda^{-1}$, these distributions have grown by about a factor of $O(10^{3})$.
\par
\begin{figure}[ht]
	\begin{minipage}{0.45\textwidth}
		\includegraphics[width=\textwidth]{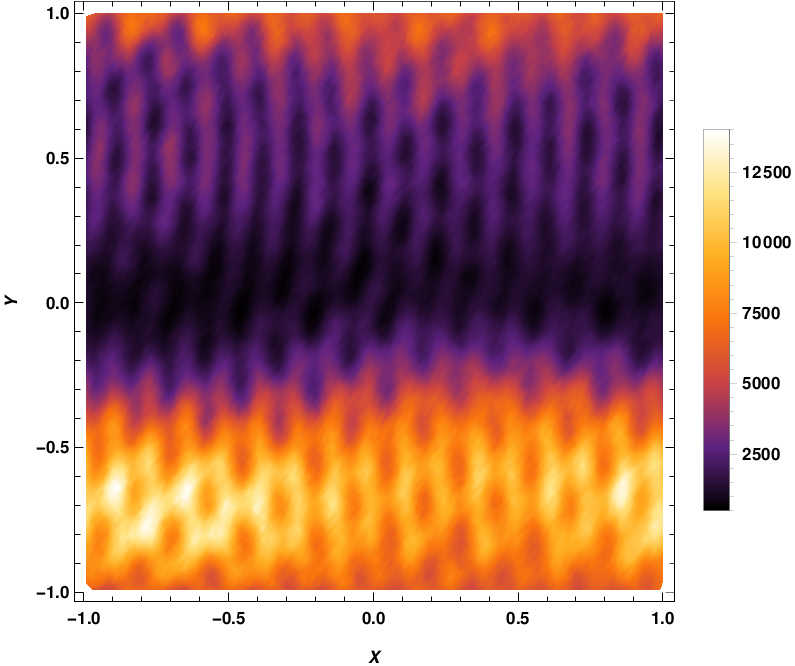}
		\caption{ ${\rm E}^2_{SU(2)}$ at t=4$\Lambda^{-1}$ for initial configuration with random fluctuations}
		\label{E2n}
	\end{minipage}
	\begin{minipage}{0.45\textwidth}
		\includegraphics[width=\textwidth]{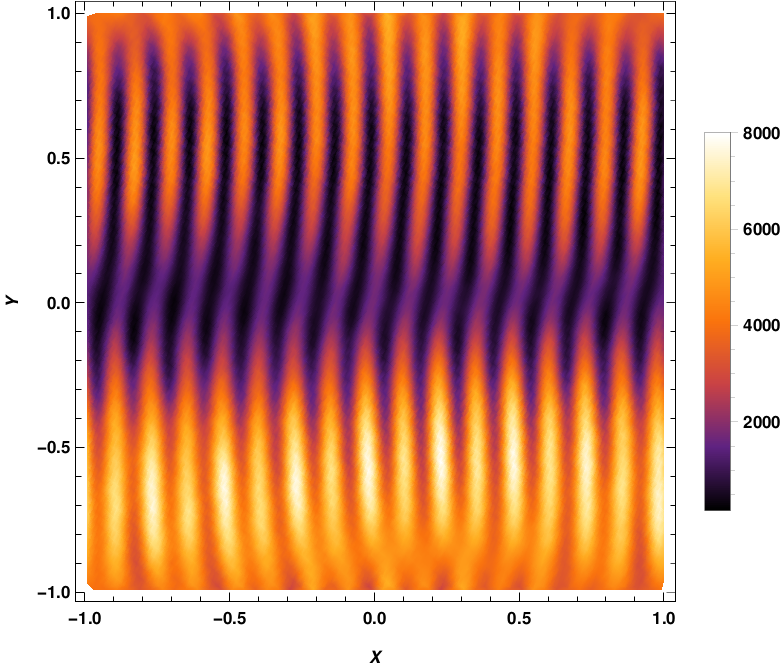}
		\caption{ ${\rm B}^2_{SU(2)}$ at t=4$\Lambda^{-1}$ for initial configuration with random fluctuations}
		\label{B2n}
	\end{minipage}
\end{figure}

 In the case of non-abelian gauge fields, we compute the gauge invariant observables,
\begin{equation}
{\rm E}^2_{SU(2)} = \sum_{i,a} F^{a,0i}F^{a,0i},~~~ {\rm B}^2_{SU(2)}= \sum_{a, i\ne j} F^{a,ij}F^{a,ij}
\end{equation}
where spatial ($i,j$) and color indices are summed over.
The results for ${\rm E}^2_{SU(2)}$ and ${\rm B}^2_{SU(2)}$ at $t=4\Lambda^{-1}$ are shown in Fig.\ref{E2n} and Fig.\ref{B2n}.  At initial time, the distributions of ${\rm E}^2_{SU(2)}$ and ${\rm B}^2_{SU(2)}$ are of the order of $O(10^{-1})\Lambda$ and $O(1)\Lambda$ respectively.
In time duration, $4\Lambda^{-1}$, their distributions grow by a factor of $O(10^4)$ and $O(10^3)$ respectively, due to parametric resonance. 

\subsection{Parametric resonance: $U(1)$ vs $SU(2)$}

\begin{figure}[ht]
	\begin{minipage}{0.45\textwidth}
	\includegraphics[width=0.95\textwidth]{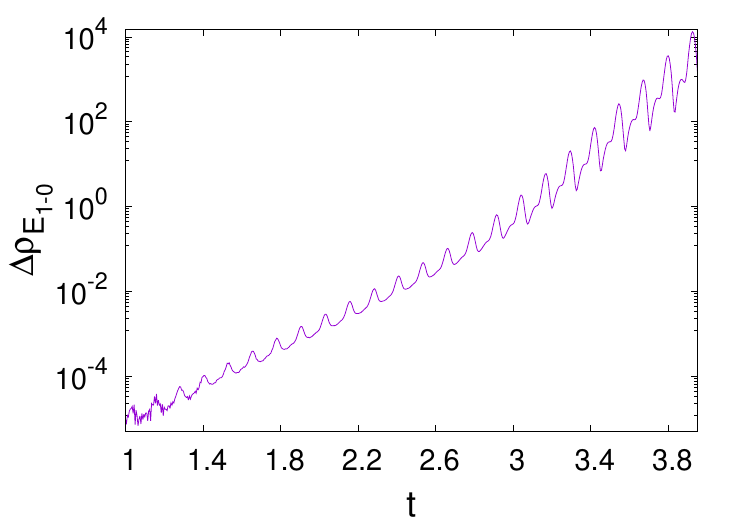}
	\caption{$\Delta\rho_{E_{1-0}}$ for random fluctuations.}
	\label{edu1su2}	
	\end{minipage}
   \begin{minipage}{0.5\textwidth}
	\includegraphics[width=0.95\textwidth]{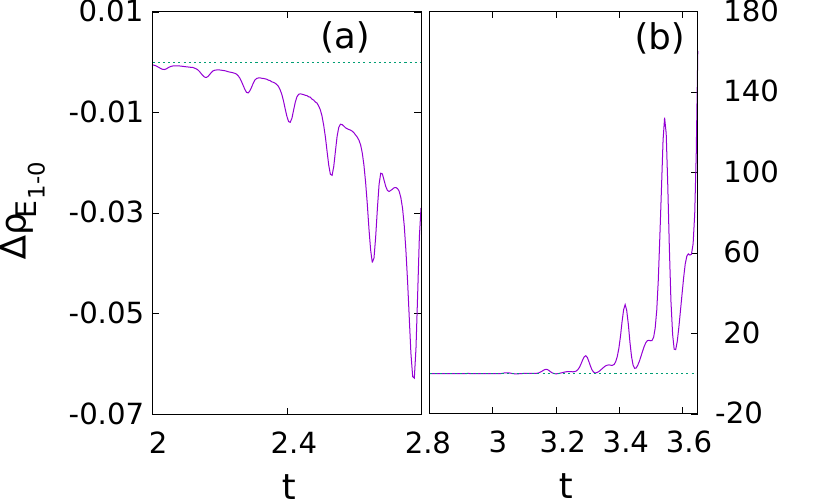}
	\caption{$\Delta\rho_{E_{1-0}}$ for random fluctuations.}
	\label{ed2}	
	\end{minipage}
\end{figure}

To study the effect of interaction in $SU(2)$ gauge theory, we compute the spatial average of energy density, $\rho_E$ for $g=0,1$. 
We considered ten different random initial configurations and evolved them with both $g=1$ and $g=0$. In half of these initial configurations, in the non-linear regime, self-interaction resulted in higher growth of energy density compared to the $g=0$ case. In Fig.\ref{edu1su2} we show the difference in the energy density between $g=1$ and $g=0$, i.e., $\Delta\rho_{E_{1-0}}$, for one of these cases. $\Delta\rho_{E_{1-0}}$ starts out very small during initial times, fluctuating around zero, which seems to be due to the presence of large momentum modes in the initial configuration. The behaviour of $\Delta\rho_{E_{1-0}}$ for other configurations is similar at initial times, but subsequently, the trough
of the fluctuations in  $\Delta\rho_{E_{1-0}}$ goes further below zero with time. In some cases, the entire curve remains below zero up to some time, as shown in Fig.\ref{ed2}$(a)$. For all the ten cases, $\Delta\rho_{E_{1-0}}$ remains positive and increases exponentially for larger times, as seen in Fig.\ref{edu1su2} and \ref{ed2}$(b)$. Similar to N-O instability, we observe an increase in $\rho_E$ with the increase in coupling constant. These results suggest that non-linear dynamics eventually enhance the growth of fluctuations. A detailed study of the power distribution in momentum space will provide a better understanding, which we plan to carry out in the future.

\begin{figure}[ht]
	\begin{minipage}{0.45\textwidth}		\includegraphics[width=0.95\textwidth]{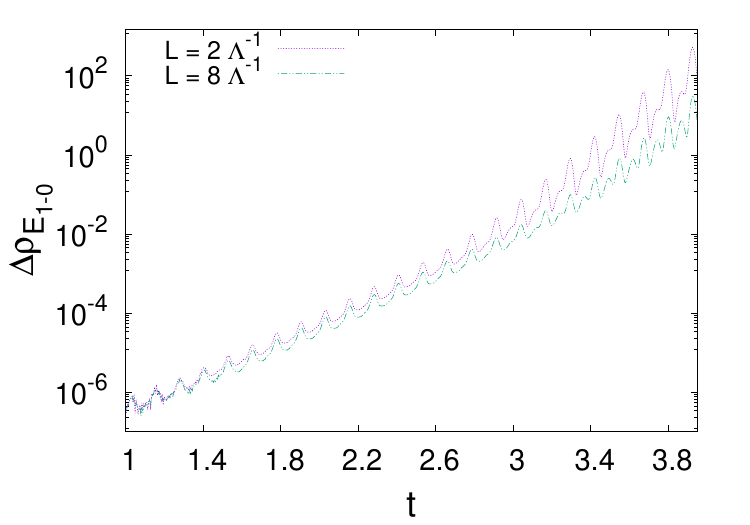}
		\caption{$\Delta\rho_{E_{1-0}}$ for $L=8\Lambda^{-1}$ and scaled $L=2\Lambda^{-1}$ results}
		\label{l7}	
	\end{minipage}
	\begin{minipage}{0.45\textwidth}
		\includegraphics[width=0.95\textwidth]{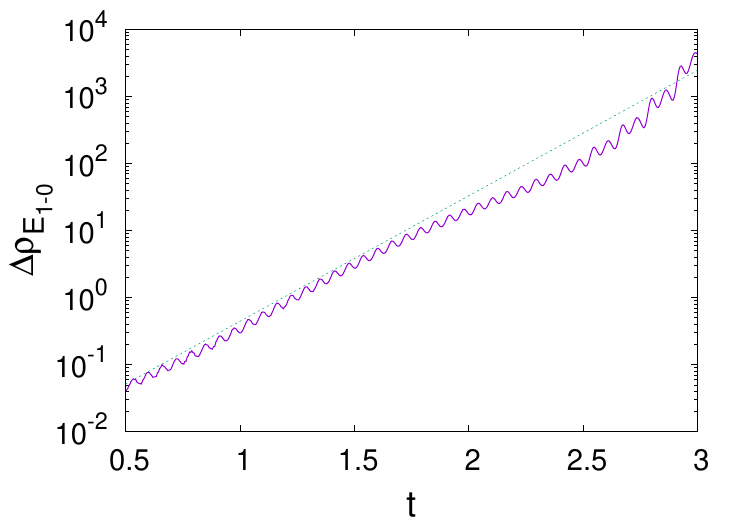}
		\caption{$\Delta\rho_{E_{1-0}}$ for $L=8\Lambda^{-1}$ with random fluctuations.}
		\label{l8}	
	\end{minipage}
\end{figure}

With time evolution, the growing field perturbations will propagate in space, as shown later in Fig.\ref{rho1_u12} and Fig.\ref{rho2_su22}. Since we are
using periodic boundary conditions, these perturbations will cross the boundary and return to the region of origin. Thus, the resonant growth is reliable up to the time the perturbations take to cross the entire lattice, i.e., $2\Lambda^{-1}$ for the simulations presented above. Until up to $t=2\Lambda^{-1}$ we clearly see resonant growth of the fluctuations. To ascertain that the evolution beyond this
time is not a numerical artifact, we consider larger lattices for the same lattice spacing. To remove the possibility of boundary effects affecting the growth, we take the same configuration as of Fig.\ref{edu1su2} at the center of the lattice of size $L=8\Lambda^{-1}$. The field is set to zero over the rest of the lattice. As a result, the distribution of the resonant modes for $L=2\Lambda^{-1}$ and $L=8\Lambda^{-1}$ are similar apart from an overall factor. Since the evolution is dominated by the resonant modes, it is expected that the growth of fluctuations in both cases will be similar. The time evolution of  $\Delta\rho_{E_{1-0}}$ is shown in Fig.\ref{l7}. We see resonant growth in $\rho_E$ for both $g=0,1$ for $L=8\Lambda^{-1}$. For comparison, we include the results for $L=2\Lambda^{-1}$ scaled by a factor $1/25$.

We also consider a different random initial configuration for $L=8\Lambda^{-1}$ lattice. Note that since the initial configuration is random, it is difficult to maintain the same distribution of spatial modes, as in $L=2\Lambda^{-1}$. The same is true for
the relative strength of the different resonant modes. In the simulation, we consider a slightly larger amplitude for the fluctuations, i.e., $\gamma=0.08\Lambda$. Fig.\ref{l8} shows $\Delta\rho_{E_{1-0}}$ in time. The dashed straight line is plotted for reference. If the range of fluctuations is kept the same as before, one observes only damping similar to Fig.\ref{ed2}$(a)$. Whereas Fig.\ref{l8} shows positive and increasing behaviour of $\Delta\rho_{E_{1-0}}$ as seen in Fig.\ref{edu1su2}. Note that the damping effects, such as in Fig.\ref{ed2}$(a)$, can also be seen relative to the reference line. The results shown in Fig.\ref{l7} and Fig.\ref{l8} suggest that perturbative feedback effects, due to periodic boundary conditions, are not significant.
	
\begin{figure}[ht]
	\begin{minipage}{0.45\textwidth}
		\includegraphics[width=0.95\textwidth]{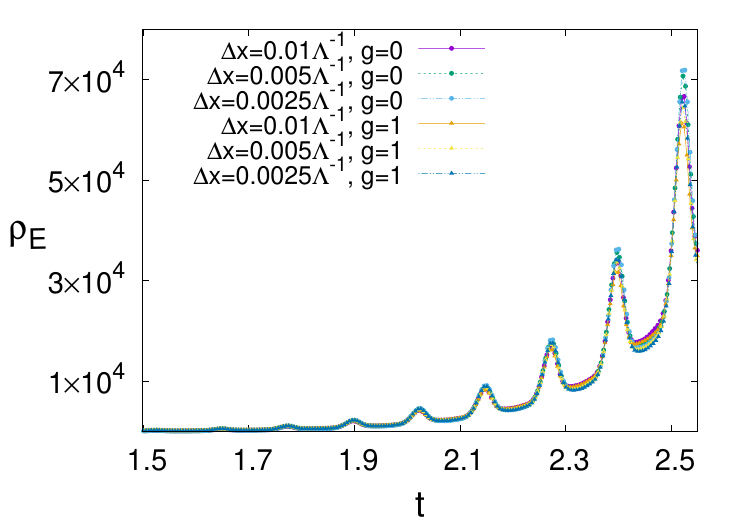}
		\caption{$\rho_E$ vs $t$ in continuum.}
		\label{edom16k8}	
	\end{minipage}
	\begin{minipage}{0.45\textwidth}
		\includegraphics[width=0.95\textwidth]{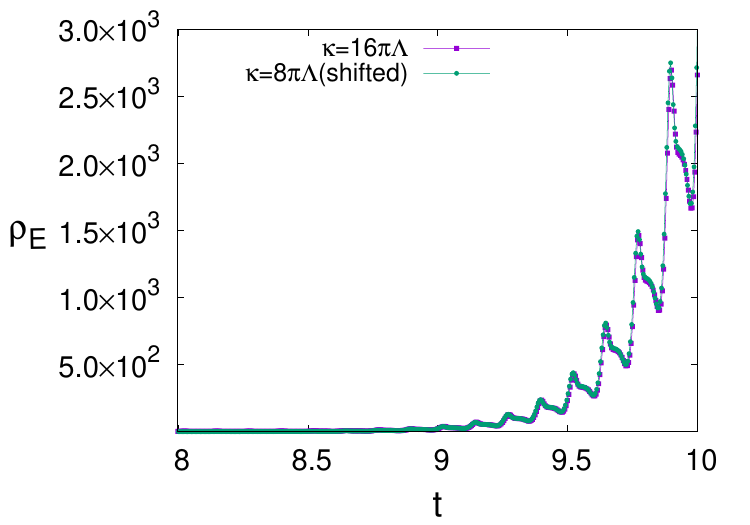}
		\caption{$\rho_E$ vs $t$ for $k_x=\omega$.}
		\label{edom16k16}	
	\end{minipage}
\end{figure}

In the above simulations, the observed resonant modes agree with those obtained analytically in section II-A. However, it is important
to check that numerical errors due to the discretisation of space and time do not affect the evolution of fields. For this purpose, we evolved an 
initial configuration for different values of $\Delta x$, i.e.,  $\Delta x=0.01\Lambda^{-1}, 0.005\Lambda^{-1}$, and $0.0025\Lambda^{-1}$. 
To maintain the same amplitude of the resonant mode at the initial time, it is necessary that the initial configuration is regular rather than a random field configuration like in the above studies. For simplicity, the initial configuration is considered to have a single resonant momentum mode along the `x' direction, e.g; $A^a_\mu=\gamma sin(\kappa^a_\mu x)$.  $\kappa^1_\mu\equiv(\omega/2,\omega/2,\omega/2,\omega/2)$, $\kappa^2_\mu\equiv(-\omega/2,-\omega,\omega,0)$ and $\kappa^3_\mu\equiv(0,\omega,-\omega,-\omega/2)$ with $\omega=16\pi\Lambda$. This particular configuration choice is made as it satisfies the periodic boundary condition, i.e., $A^a_\mu(x=0)=A^a_\mu(x=L)$, with $L=2\Lambda^{-1}$. Note that $\kappa_0$ has nothing to do with temporal variations of the gauge field, but it represents the momentum mode of $A^a_0$ in `x' direction. 
The evolutions of $\rho_E$ are shown in Fig.\ref{edom16k8}, for three different choices of lattice spacing. At late
times the evolution of the difference in energy density for $\Delta x = 0.005\Lambda^{-1}$ and $\Delta x = 0.0025\Lambda^{-1}$ is much
smaller compared to that for  $\Delta x = 0.01\Lambda^{-1}$ and  $\Delta x = 0.005\Lambda^{-1}$. Further, the time evolution of this configuration shows that the energy density does not vary significantly with the gauge coupling $g$. However, the evolution of the energy density of each color, $\rho_E^a$, depends on $g$. This suggests that there is an exchange of energy density between different colors. This exchange of energy density persists for higher $g$ we studied in our simulations. 

The above results show that discretisation errors do not significantly affect the evolution of the fields. However, it is
not practical to specify a field configuration with only one mode on a discrete lattice. A continuous function discretised on the lattice will
always have a non-zero overlap with other modes. In a follow-up simulation, we considered, the initial configuration with only one color, $A^a_\mu=\delta_{a,1}\gamma sin(\kappa^a_\mu x)$, i.e., the abelian case, with all components being $\kappa^1_\mu\equiv(\kappa,\kappa,\kappa,\kappa)$ for $\kappa=8\pi\Lambda~(\omega/2)$ and $16\pi\Lambda~(\omega)$. The evolution of $\rho_E$ for $\kappa=16\pi\Lambda$ is shown in Fig.\ref{edom16k16}. $\rho_E$ is small initially, up to $t\sim 8\Lambda^{-1}$, but subsequent evolution is similar to $\kappa=8\pi\Lambda$ case. For comparison, $\rho_E$ corresponding to $\kappa=8\pi\Lambda$ is shifted in time(here by $\simeq7.625\Lambda^{-1}$). This indicates that the $\kappa=8\pi\Lambda$ mode is initially present, due to the discretization of the $\kappa=16\pi\Lambda$ mode. Though the amplitude is small, eventually, it dominates due to parametric resonance. This result suggests that the lowest resonant mode always dominates the evolution. In the following, we discuss our numerical simulations in $3+1$ dimensions.
\par 
In this section, so far, the fields are considered to be independent of $z$. However, since the space-time oscillations are propagating along the $z-$direction, the metric depends on $z$ along with time $t$. Thus, in $3+1$ dimensions, even in the linear regime, there is coupling between momentum modes that differ by $\Delta k_z=\pm\omega$. In the case of $SU(2)$, the $k_z$ modes will also get excited due to non-linear interactions. 

\par
\begin{figure}[ht]
	\begin{minipage}{0.45\textwidth}
		\includegraphics[width=0.95\textwidth]{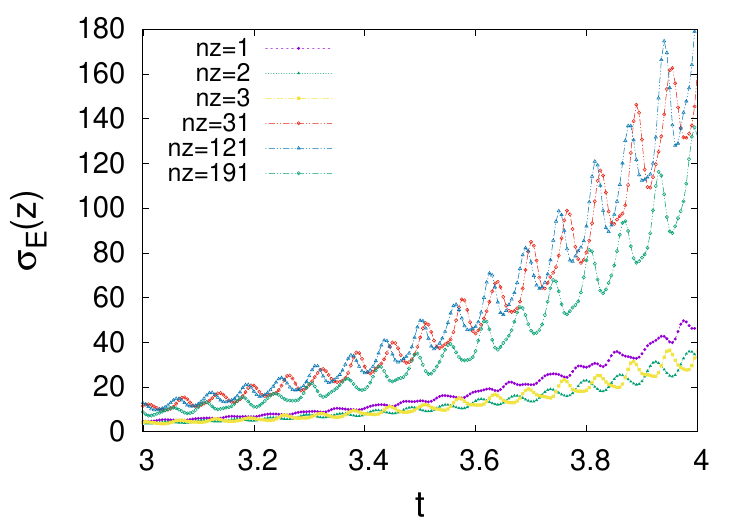}
		\caption{$\sigma_{E}(z)$ for different $xy-$planes.}
		\label{edzvar}	
	\end{minipage}
	\begin{minipage}{0.45\textwidth}
		\includegraphics[width=0.95\textwidth]{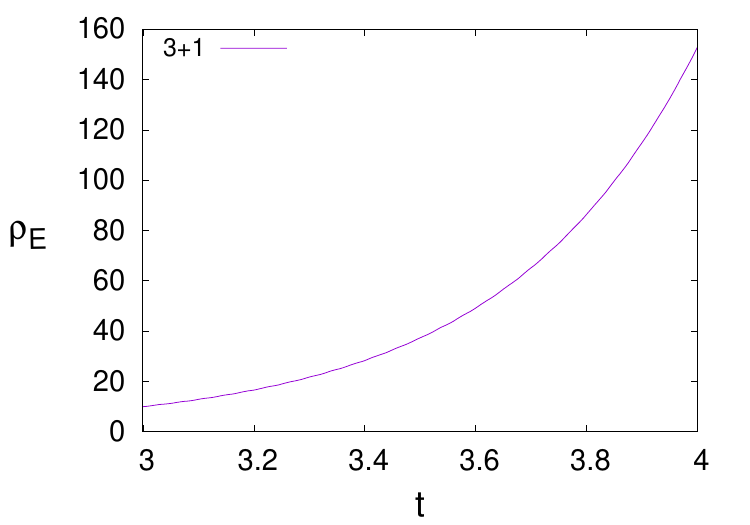}
		\caption{$\rho_E$ vs $t$ for $3+1$ dimensions.}
		\label{edkzr}	
	\end{minipage}
\end{figure}

 
In order to understand the effect of $k_z$ modes, we carry out simulations in $3+1$ dimensions for both $U(1)$ and $SU(2)$. We consider the $3-$dimensional cubic system with size $(5\Lambda^{-1})$ having same range of fluctuations, as in $2+1$ dimensional simulations, $[-\gamma,\gamma]$ with $\gamma=0.005\Lambda$. Here considered lattice constants are $\Delta x=\Delta y=2\Delta z=0.05\Lambda^{-1}$. The discrete time step is taken to be $\Delta t=0.005\Lambda^{-1}$. In Fig.\ref{edzvar} we show the time evolution of the surface energy density, i.e $\sigma_E(z,t)$, for three neighboring planes for $nz=1,2,3$ and also for $nz=31,121,191$. One can see that the time dependence of $\sigma_E$ is similar to that of the $2+1$ case, i.e. exponentially increases in time with an additional oscillatory component. The growth of fluctuations in $2+1$ is found to be stronger than in $3+1$. It is found that the exponent growth($\eta$) of $\sigma_E(z,t)$ varies mildly with $z$, i.e $\eta=2.8\pm 0.4$. As a consequence, the average energy density $\rho_E$ also grows with a similar exponent, shown in Fig.\ref{edkzr}. Note that the evolution shown here is for time up to, $t=4\Lambda^{-1}$, which is smaller compared to the size of the system, to exclude possible effects of the periodic boundary conditions.

\begin{figure}[ht]
	\begin{minipage}{0.45\textwidth}
		\includegraphics[width=0.95\textwidth]{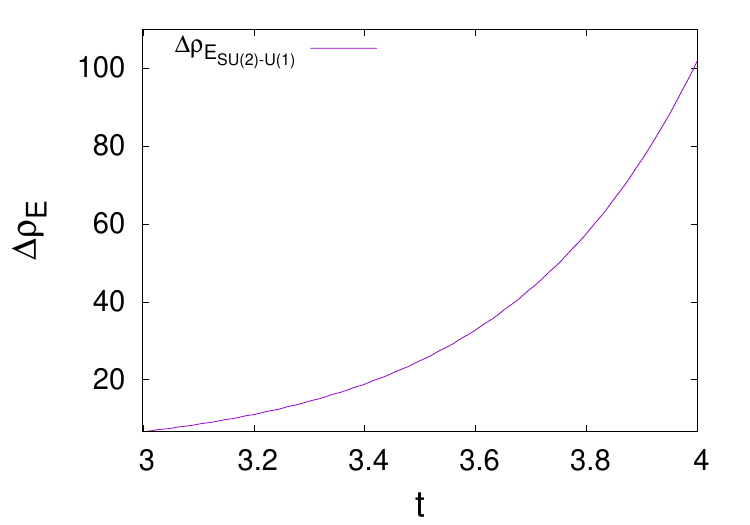}
		\caption{$\Delta\rho_{E}$ between $SU(2)$ and $U(1)$.}
		\label{edkzrsu2u1}	
	\end{minipage}
\end{figure}
In Fig.\ref{edkzrsu2u1} we plot the energy difference between the energy density between $SU(2)$ and $U(1)$. The results show that the growth of energy density in $SU(2)$ still dominates over $U(1)$ due to the color factor.

\subsection{CP violations due to space-time oscillations}

\begin{figure}[ht]
	\begin{minipage}{0.45\textwidth}
		\includegraphics[width=\textwidth]{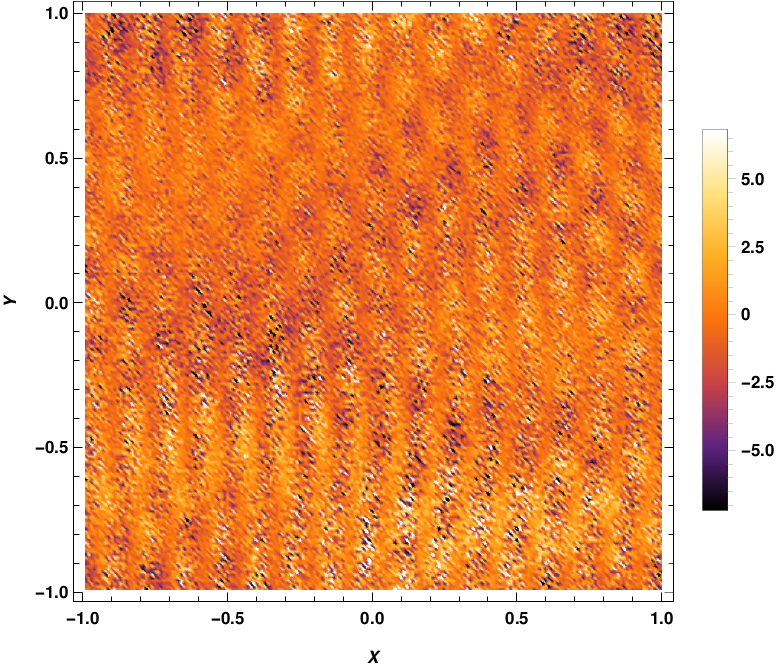}
		\caption{ $\rho_1$ at $t=3\Lambda^{-1}$ for  initial configuration with random fluctuations}
		\label{rho13}
	\end{minipage}
	\begin{minipage}{0.45\textwidth}
		\includegraphics[width=\textwidth]{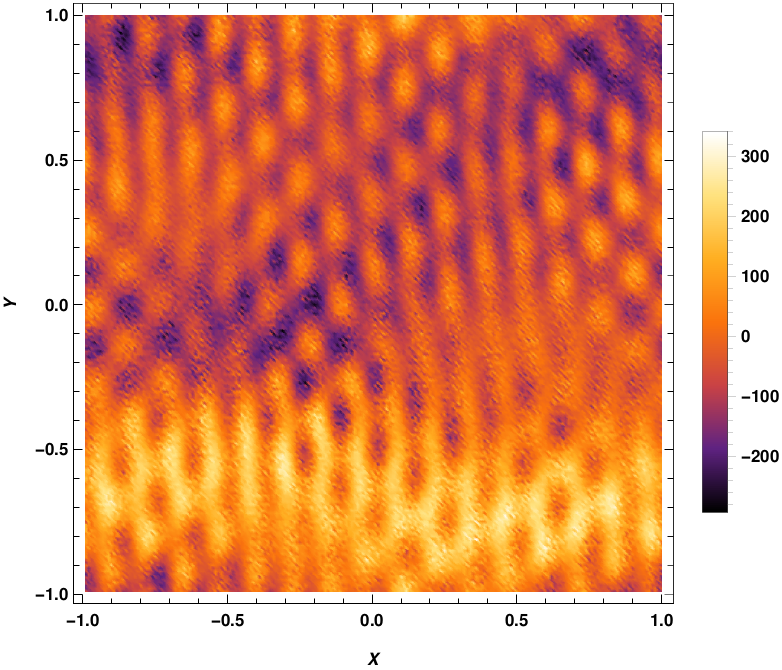}
		\caption{ $\rho_1$ at $t=4\Lambda^{-1}$ for initial configuration with random fluctuations}
		\label{rho14}
	\end{minipage}
\end{figure}

\begin{figure}[ht]
	\begin{minipage}{0.45\textwidth}
		\includegraphics[width=\textwidth]{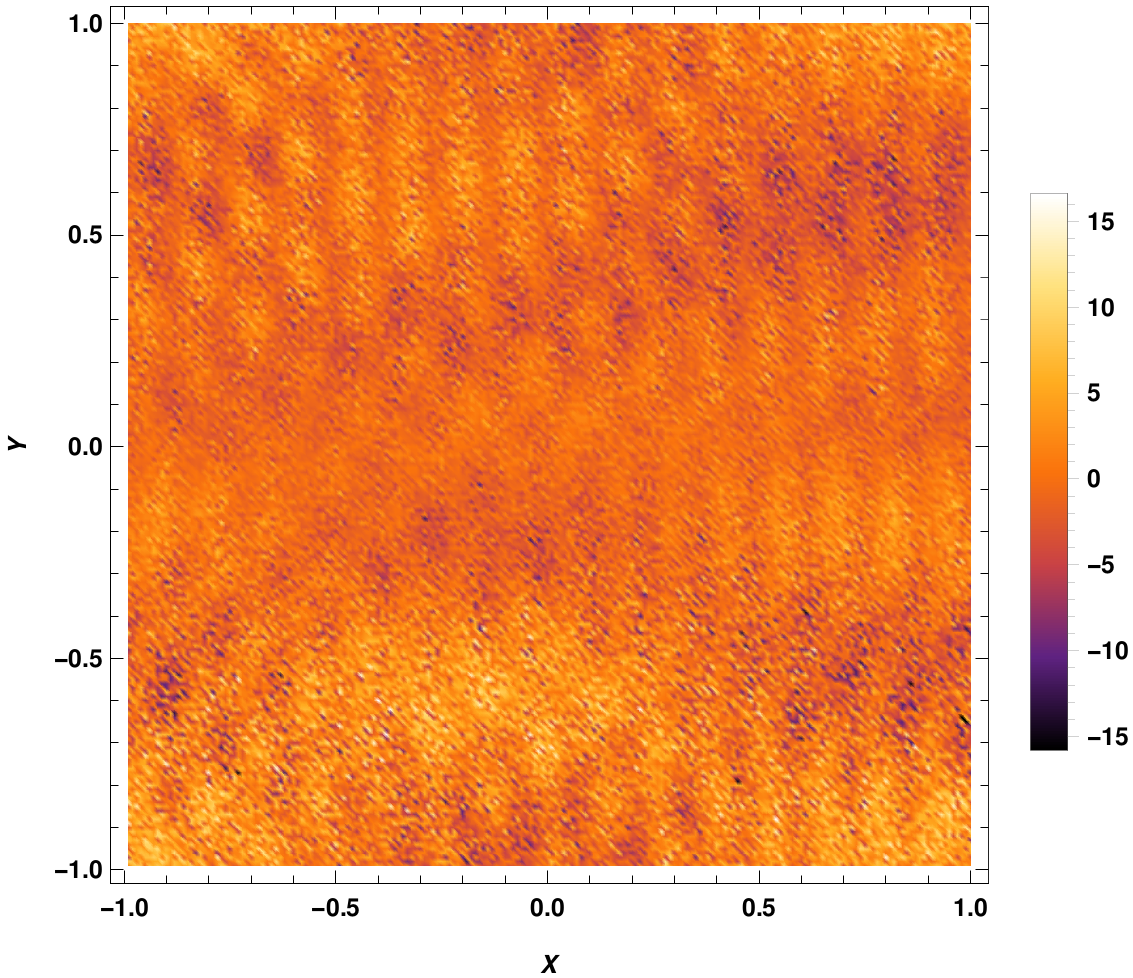}
		\caption{ $\rho_2$ at $t=3\Lambda^{-1}$ for  initial configuration with random fluctuations}
		\label{rho21}
	\end{minipage}
	\begin{minipage}{0.45\textwidth}
		\includegraphics[width=\textwidth]{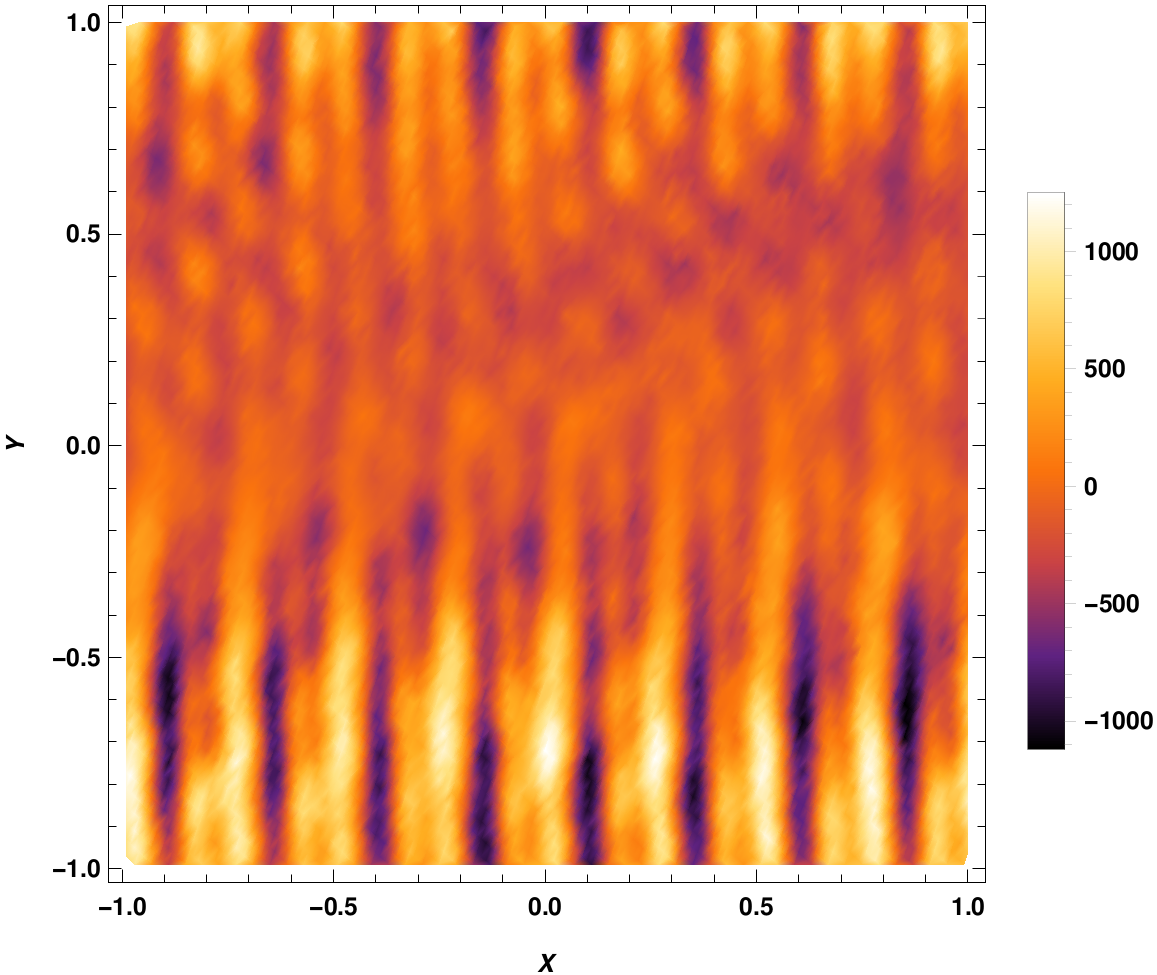}
		\caption{ $\rho_2$ at $t=4\Lambda^{-1}$ for initial configuration with random fluctuations}
		\label{rho22}
	\end{minipage}
\end{figure}

In the following, we present our results for the evolution of the distribution $\rho_1={\bf E}\cdot{\bf B}$ and $\rho_2={\bf E}^a\cdot{\bf B}^a$ in $U(1)$ and $SU(2)$ respectively. The initial value of $\rho_1$ is about $O(10^{-1})$. In Fig.\ref{rho13} and Fig.\ref{rho14}, $\rho_1$ is shown at time $t=3\Lambda^{-1}$ and $t=4\Lambda^{-1}$. These figures show that over the lattice $\rho_1$ is distributed around zero. Within a time span of $\Lambda^{-1}$ the highest (lowest) value increase (decrease) by a factor of $\sim 20$.\\
The figures, Fig.\ref{rho21} and Fig.\ref{rho22}, show $\rho_2$ at $t=3\Lambda^{-1}$ and $t=4\Lambda^{-1}$ respectively. Initially, the distribution of $\rho_2$ is within $[-0.5,0.5]\Lambda^4$. This range grows to $\sim[-17,18]\Lambda^4$ and $\sim[-1200,1300]\Lambda^4$, at $3\Lambda^{-1}$ and $4\Lambda^{-1}$ respectively. 
\par  The initial distribution of $\rho_1$ and $\rho_2$, in the above simulations, are non-zero. We have also checked evolution of configurations for which $\rho_1$ and $\rho_2$ vanish at initial times.  An initial condition with vanishing $\rho_1$ and $\rho_2$ is difficult to achieve, with random fluctuations of the gauge fields. Thus, we considered gauge fields such that $\rho_1=0=\rho_2$, while the field strength tensors were non-zero in a localized region. Note that, the initial configuration needs to have non-zero (preferably resonant) modes, in order to couple to the space-time oscillations. Hence, we considered a gaussian form for the gauge fields for $U(1)$, i.e.,
\begin{equation}
{\bf A}(t=0)=-\beta\Delta t e^{-\alpha (x^2+y^2)}\hat{k},~ A_0(t=0)=0;~~ {\bf A}(t=\Delta t)=0,~ A_0(t=\Delta t)=0
\label{inc}
\end{equation} 
$\alpha$ is taken to be the time period of one space-time oscillation. This makes sure that the initial configuration has non-zero resonant mode.
\begin{figure}[ht]
	\begin{minipage}{0.45\textwidth}
		\includegraphics[width=\textwidth]{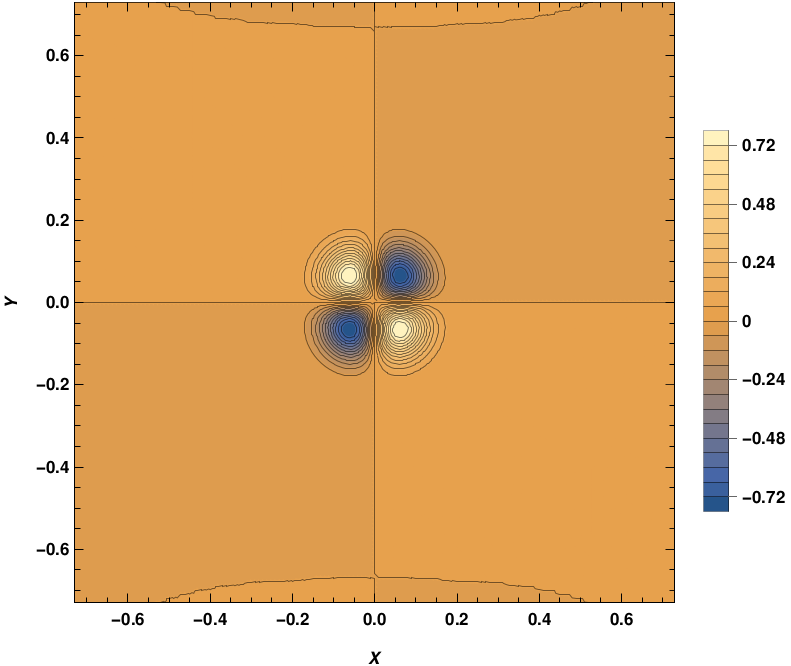}
		\caption{ $\rho_1$ at $t=0.10\Lambda^{-1}$}
		\label{rho1_u11}
	\end{minipage}
	\begin{minipage}{0.45\textwidth}
		\includegraphics[width=\textwidth]{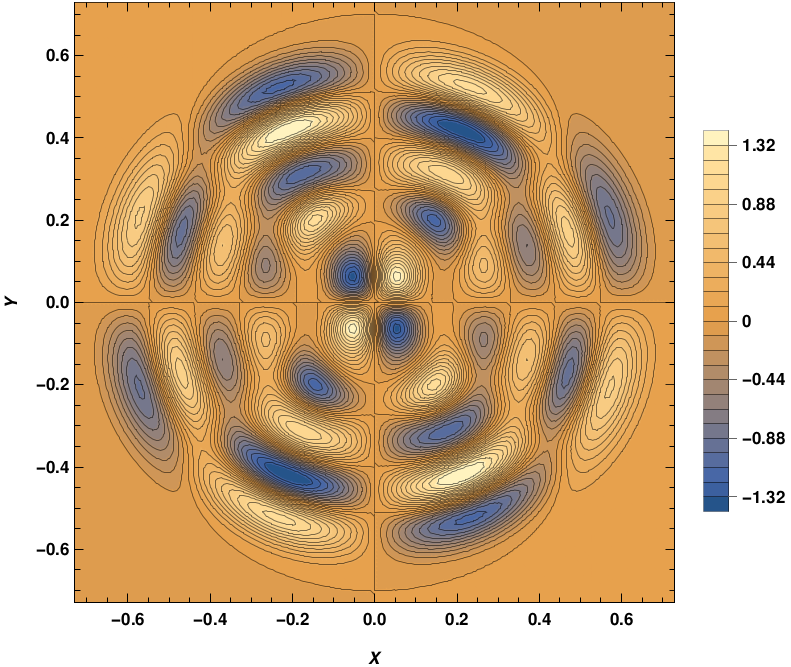}
		\caption{ $\rho_1$ at $t=0.60\Lambda^{-1}$}
		\label{rho1_u12}
	\end{minipage}

	\begin{minipage}{0.45\textwidth}
		\includegraphics[width=\textwidth]{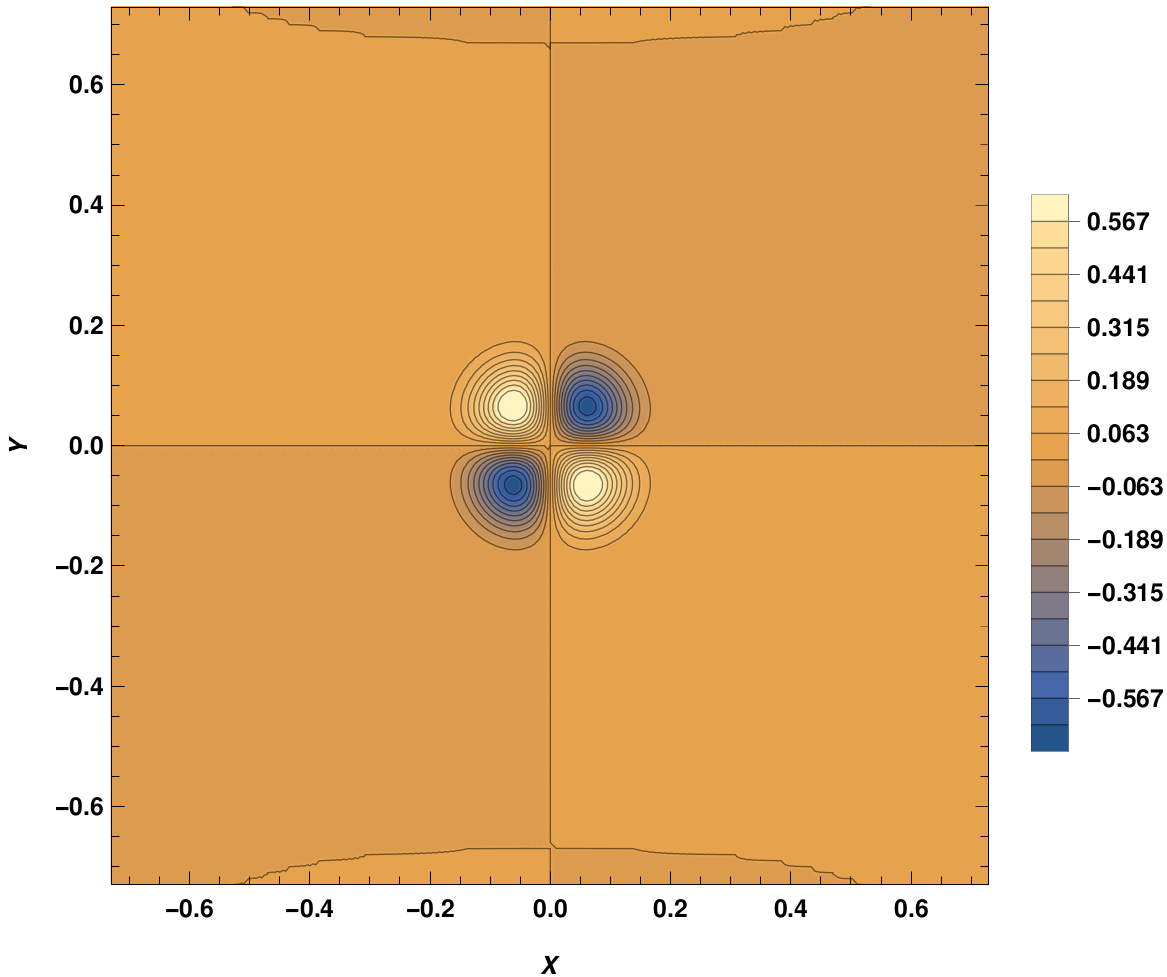}
		\caption{ $\rho_2$ at $t=0.10\Lambda^{-1}$}
		\label{rho2_su21}
	\end{minipage}
		\begin{minipage}{0.45\textwidth}
		\includegraphics[width=\textwidth]{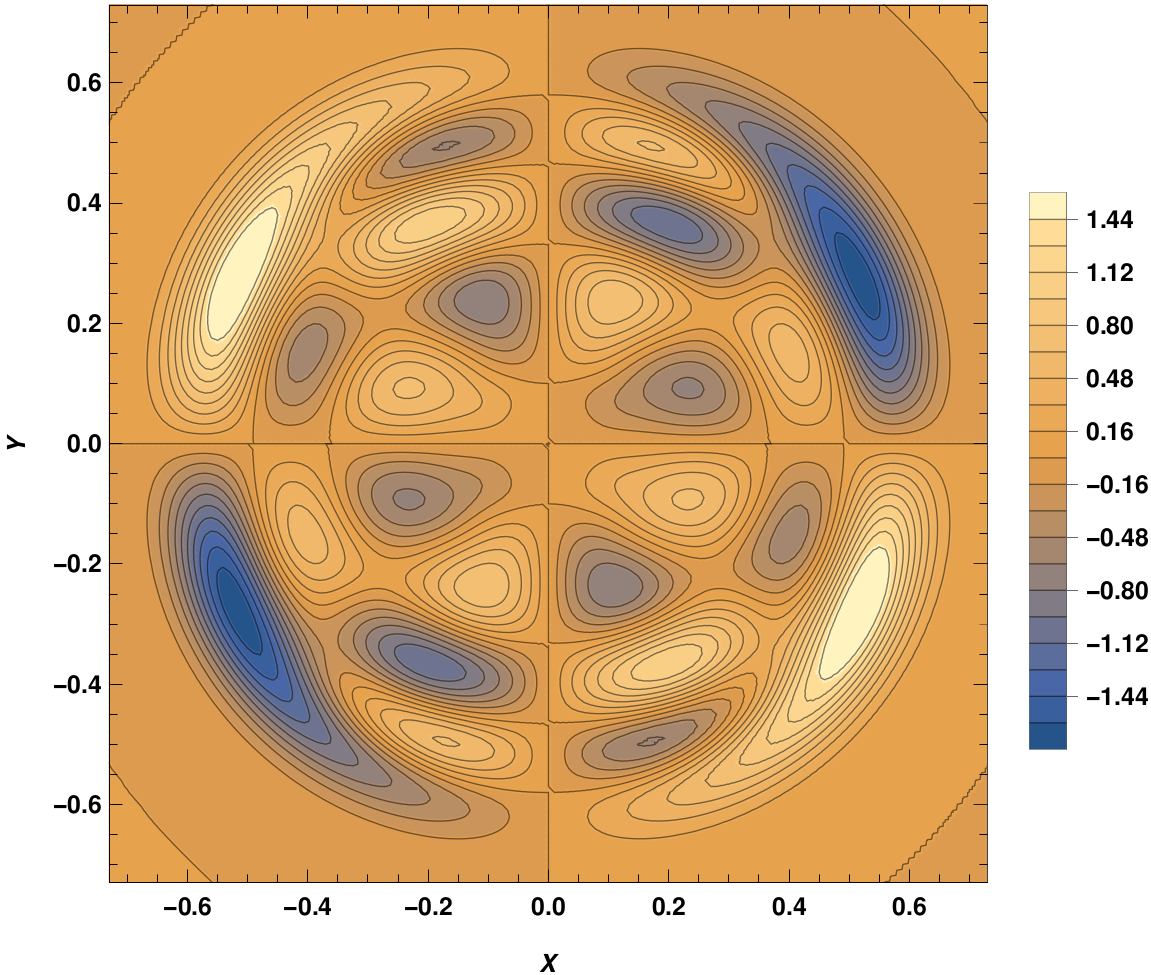}
		\caption{ $\rho_2$ at $t=0.60\Lambda^{-1}$}
		\label{rho2_su22}
	\end{minipage}
\end{figure}
\par In $SU(2)$, when two of the three color fields are zero, the evolution of the non-zero field is
similar to the $U(1)$ case. Hence, we consider configurations for which the gauge fields are non-zero for at least two colors. Also, non-linear evolution requires that field strength and coupling constant should be large enough, e.g; here $\beta\Delta t \sim 0.1\Lambda$ and $g=1.5$. Note that even with smaller $\beta$ values the evolution will be non-linear at later times. The localised configurations, Eq.\ref{inc}, spread quickly during evolution. Once the fluctuations reach the boundary of the lattice, they lead to systematic errors in evolution. 
\par The figures, Fig.\ref{rho1_u11} and Fig.\ref{rho1_u12}, show the distributions of $\rho_1$ at $t=0.10\Lambda^{-1}$ and $t=0.60\Lambda^{-1}$ respectively. The results for $\rho_2$ are shown in Fig.\ref{rho2_su21} and \ref{rho2_su22} for $t=0.10\Lambda^{-1}$ and $t=0.60\Lambda^{-1}$
respectively. These results clearly show that $\rho_1$ and $\rho_2$ can be excited starting from zero through
space-time oscillations. Further, the effect of
space-time oscillations is larger on $\rho_2$ due to non-linear interaction between the gauge fields. Note that, the persistent exponential increase observed in CP both for $U(1)$ and $SU(2)$ 
will saturate due to the turbulent cascade mechanism studied in ref.\cite{Mace:2019cqo}.
\section{Discussion and conclusion}

We have studied the evolution of abelian $U(1)$ and non-abelian $SU(2)$ gauge fields in the presence of monochromatic space-time oscillations, using analytical and numerical simulations. Our results show that space-time oscillations induce parametric resonance in the modes of gauge fields.  The fields' modes undergoing parametric resonance are obtained from field equations for $U(1)$ and that for $SU(2)$ in linear approximation in Fourier space. We also carry out numerical simulations in $2+1$ and $3+1$ dimensions. These simulations are useful for studying the evolution of localized initial field configurations in $U(1)$. In the case of $SU(2)$ gauge fields, they are essential to study field dynamics beyond the linear regime. The numerical simulations also enable us to compare the dynamical evolution in the $U(1)$ and $SU(2)$ gauge theory, in other words, effect of gauge coupling.

\par In $2+1$ dimensions, the process of the parametric resonance of the gauge field modes generates large fluctuations in physical observables. These observables include $F_{\mu\nu}\tilde{F}^{\mu\nu}$ and $F^a_{\mu\nu}\tilde{F}^{a\mu\nu}$ in $U(1)$ and $SU(2)$ gauge theory respectively, which break the $CP-$symmetry. Apart from the color factor, the evolution of gauge fields is similar in both theories, when the fluctuations are small. At later times when the gauge field fluctuations grow, the effects of non-linear interactions in $SU(2)$ become important. The non-linear interactions drive larger growth in the fluctuations, beyond the conventional exponential growth expected in linear theories. The dynamics of the non-abelian gauge fields were found to be similar to that in the case of N-O instability due to chromoelectric and magnetic fields. Our results also show that in both the $U(1)$ and $SU(2)$ cases the dynamics of the evolution are dominated by the lowest resonant mode, i.e., $k=\omega/2$ mode. In 3+1 dimensions  simulations also we see resonant growth of fluctuations. However, the exponent of growth of the energy density is found to be smaller compared to the case of $2+1$ dimensions.

\par Our numerical simulations suggest that gravitational waves also will lead to parametric resonance in gauge fields. The gravitational waves will be dampened more efficiently due to the color factor. Further, the gravitational waves can induce large fluctuations in $CP$ violating physical observables. These observables can subsequently lead to a large-scale imbalance in local chiral charge distributions. Also a large non-zero ${\bf E}\cdot{\bf B}$ can enhance production of axions. The gravitational waves, in the early Universe also have the potential to generate large-scale magnetic fields, corresponding to resonant modes. 
\acknowledgements

We thank A. P. Balachandran, Amruta Mishra for valuable discussions and suggestions.

\end{document}